# Mathematical Toolbox and its application in the Development of Laboratory Scale Vertical Axis Wind Turbine

**Aravind CV[1], Rajparthiban R[2], Rajprasad R[3],**
[1] Taylors University, Selangor Malaysia
[2]Manipal International University, Kelanajaya, Malaysia
[3]University of Nottingham, Malaysia
[1]aravindcv@ieee.org

**Grace I[4], Rozita Teymourzadeh [4], M.Norhisam[5]**
[4]UCSI University, Cheras, Malaysia
[5] University Putra Malaysia, Selangor, Malaysia

*Abstract*— **Wind turbine works with the principle of extracting energy from the wind to generate electricity. The power generated is directly proportional to the wind speed available. There are two major types of wind turbine design namely the horizontal and vertical axis wind turbine depending on the orientation of the turbine rotor and its generator. This paper deals with the design of vertical turbine due to its advantage of operating at a low wind speed over that of horizontal turbine. The analysis of change in the parameters of a vertical axis wind turbine is investigated to get the optimized way in which the rotor of the turbine is to be designed. This is done through modelling and simulation of the turbine using various parameters in the MATLAB/SIMULINK environment. A graphical user interface is created for a generic model of vertical axis wind turbine that is used to determine its parameters.**

***Keywords-Vertical Axis, SIMULINK, mechanical power, wind turbine***

## I. INTRODUCTION

Wind turbine power generation are depends on the wind speed available and the design of the turbine. The amount of power derived from the available wind resource using Horizontal Axis Wind Turbine (HAWT) are greatly on the radius of the turbine and the wind speed. The height of the rotor turbine seldom brings effect on its power generation capability. In other words, to have a HAWT turbine, the rotor of the turbine has to be placed in a manner in which the rotor blades are not obstructed from the wind.

This function makes the Vertical Axis Wind Turbine (VAWT) to have a reasonable edge to that of the HAWT since it does not have to be mounted very high. It can be mounted even on top of a building. The VAWT also has the advantage of the height of the turbine rotor playing a significant contribution to the amount of wind power generated. Another major advantage of VAWT over the HAWT is the fact that VAWT is suitable at both very low wind speed and extreme wind speeds whereas HAWT cannot be used in such situations. The analysis of VAWT in the paper is for low wind speed of less than 5*m/s*.

## II. DESIGN OF VAWT

To generate power, the turbine depends on its physical parameters, wind speed, mechanical speed of the generator and the tip speed ratio of the turbine. The average of the wind speed is given by Eq. (1)

$$V_{ave} = \frac{V_1+V_2}{2} \quad (1)$$

where $V_1$ and $V_2$ are the inlet and outlet wind speeds in m/s.
The derived kinetic energy is given by Eq. (2)

$$K_E = \frac{1}{2}mV^2 \quad (2)$$

where m is the mass of the airflow
Therefore, the power extracted is given by Eq. (3)

$$P = \frac{1}{2}m({V_1}^2 - V_2^2) \quad (3)$$

Substituting the mass in Eq.(3), the power that the rotor can extract from the wind is given in Eq.(4)

$$P_m = \frac{\rho}{4}({V_1}^2 - V_2^2)(V_1 + V_2)A \quad (4)$$

The available power from the wind is given by Eq. (5)

$$P_a = \frac{1}{2} m V_1^{\ 2} \quad (5)$$

And substituting the mass, then the available power is given Eq. (6)

$$P_a = \frac{1}{2} \rho A V_1^{\ 3} \quad (6)$$

Introducing the power coefficient $C_p$ as in Eq. (7)

$$C_p = \frac{P_m}{P_a} \quad (7)$$

Therefore, the mechanical power $P_m$ generated by the turbine is then seen as in Eq. (8)

$$P_m = \frac{1}{2} \rho A V^3 C_p \quad (8)$$

where ρ is air density (kg/m3), *A* is turbine blade area [$m^2$] $C_p$ is the power coefficient, *V* is the wind speed [m/s] [1] [4]. The equations required for the design of a vertical wind turbine is similar to that of horizontal wind turbine. The difference lies in the area of the different surface of the type used. The vertical axis wind turbine has its area as Eq. (9)

$$A = 2RH \quad (9)$$

For the horizontal turbine the area is given as in Eq. (10)

$$A = \pi R^2 \quad (10)$$

where R is the radius of the turbine and H is the height of the turbine. This goes to prove why the height of the rotor is significant for VAWT. The ratio of the blade tip speed to the wind stream, the tip speed ratio is given as in Eq.(11)

$$\lambda = \frac{R\omega}{V} \quad (11)$$

where ω is the turbine speed and *V is* the wind speed.

Based on the tip speed ratio, the power coefficient, the amount of energy that can be taken from the wind is calculated as in Eq. (12)

$$C_p(\lambda, \beta) = C_1 \left( \frac{C_2}{\lambda_i} - C_3\beta - C_4 \right) e^{-\frac{C_5}{\lambda_i}} + C_6\lambda \quad (12)$$

The coefficients $C_1$ to $C_6$ are: $C_1$ = 0.5176, $C_2$ = 116, $C_3$ = 0.4, $C_4$= 5, $C_5$ = 21 and $C_6$= 0.0068 [3], *β* is the pitch angle and is angled at which the wind hits the blades. The equations stated above have been implemented in SIMULINK and a graphical user interface to aid a user in designing their own VAWT. Fig. 1 shows the skeleton of the designed graphical user interface (GUI) whilst the corresponding modeling blocks is shown in Fig. 2.

## III. Model Analysis

The proposed model is designed with parameters as R=0.2, H=0.3, ω=73.3rad/s, Wind speed =5and N=8.The model is based partly on the Zephr VAWT and works on the principle of MAGLEV. Fig. 3 shows the structural characteristics of the VAWT designed whilst Fig.4 shows the fabricated laboratory scaled MAGLEV concept based wind turbine using CAD models developed in principles [6-8]. Using the equations stated, **λ**=2.93, $C_p$=0.0455, $T_m$=0.0057 and $P_m$ = 0.4177 where $T_m$ is the motor torque. This proves that the model present cannot generate a high-quality torque as required. This certainly will affect the power coefficient of the wind turbine. The graph of $P_m$ versus $C_p$ is derived at various turbine speeds between 1rad/s and 75 rad/s as illustrated in Fig.3.

A comparison of the experimental result versus the analytical result for the current model is shown in the Fig. 6. It is found that the analytical results yields more power as compared the experimental results because of the difference in the improper assumptions made on the analytical. Hence a better analysis on the parameter has to be analysed. In order to improve the design of the wind turbine the factors influencing are the power co-efficient *Cp*, the area of the contact surface and the wind speed. The wind speed of the analysis is kept at a constant value and hence the approach for improvement of the design is by varying $C_p$ and by the area of the turbine. To observe the performance of VAWT, the turbine characteristics are simulated for various parameters such as the different radius, the height, the turbine speed, the tip speed ratio while keeping the wind speed at a constant value of 2m/s.

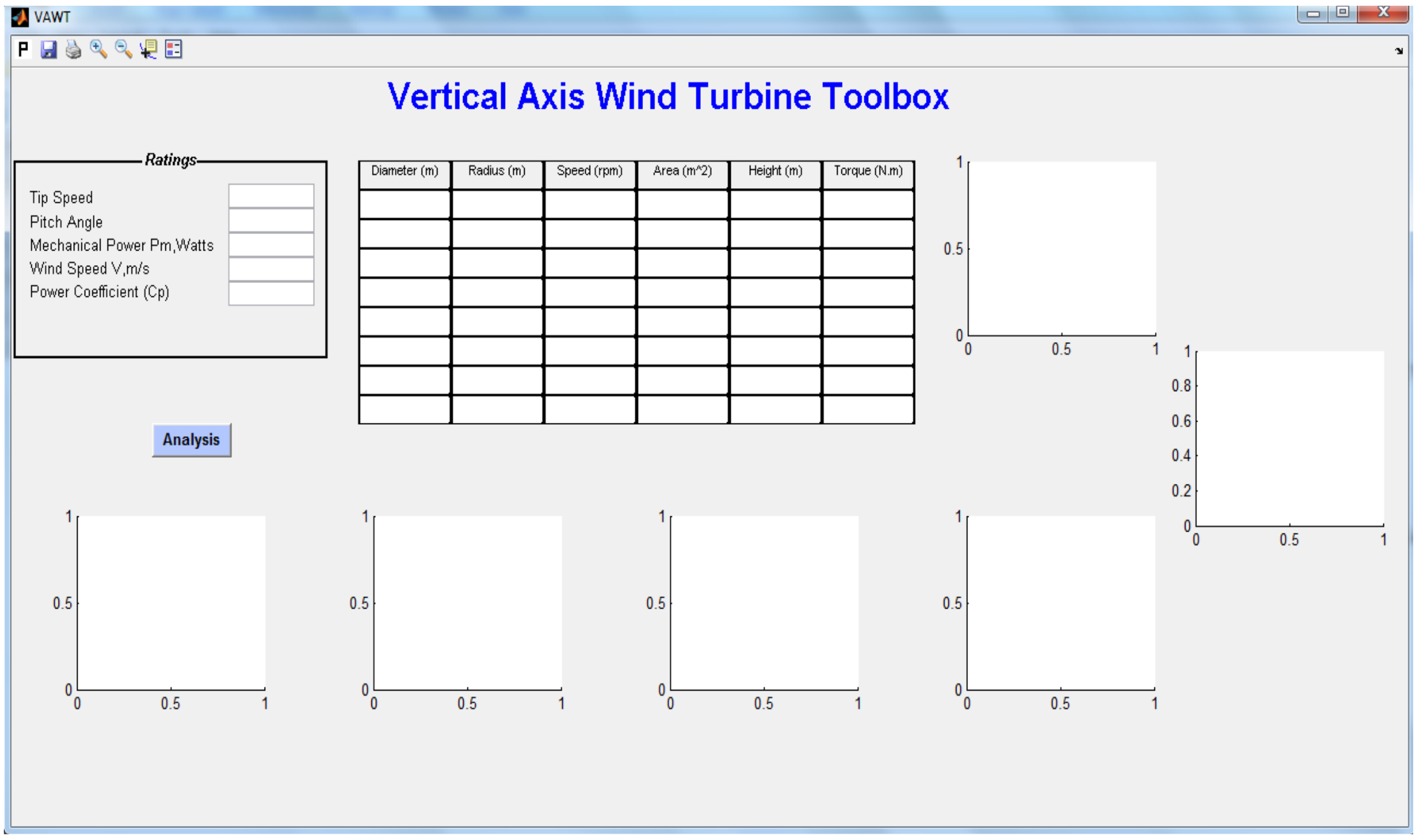

**Figure 1: Turbine SIMULINK model**

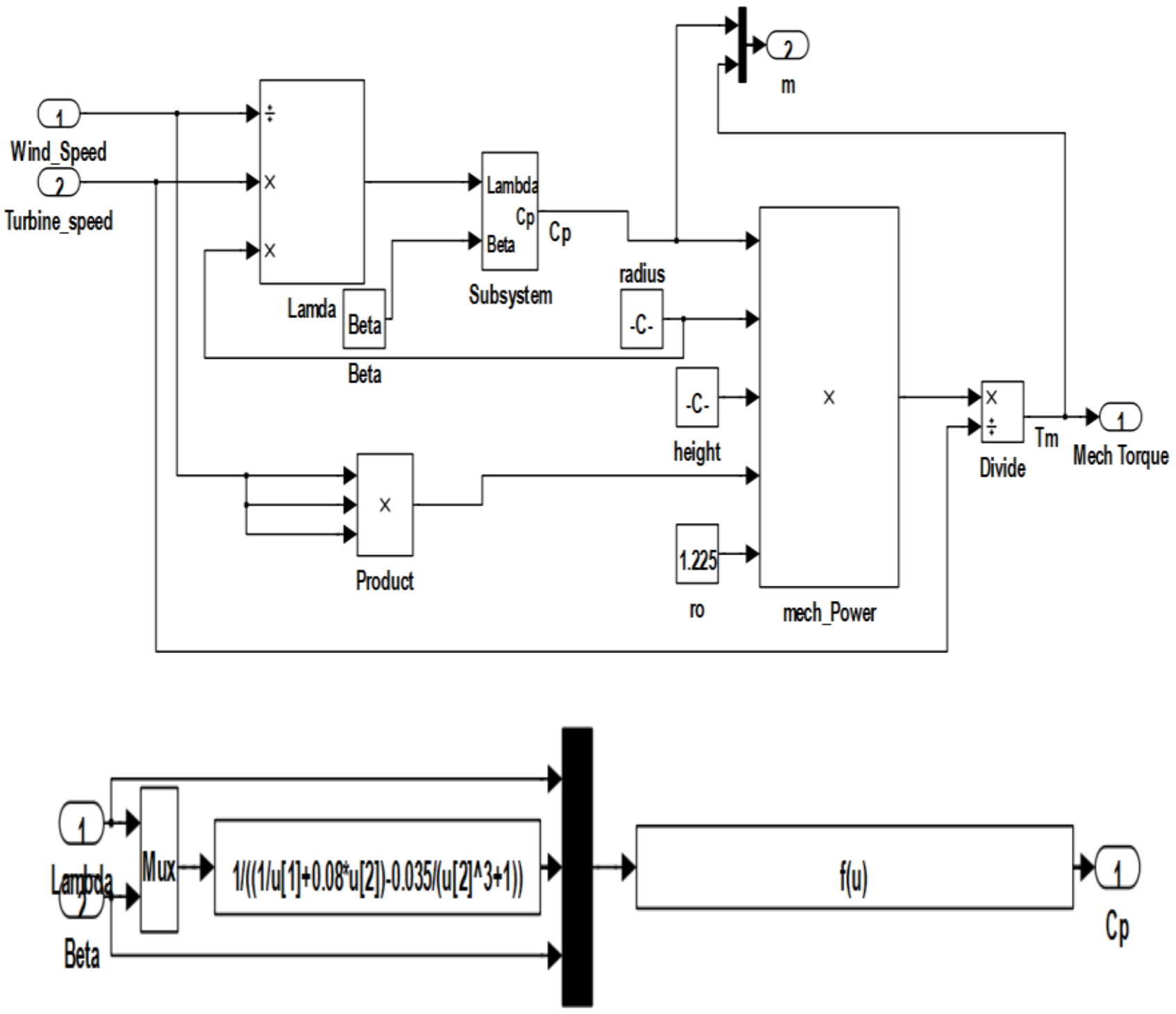

**Figure 2: Modeling blocks**

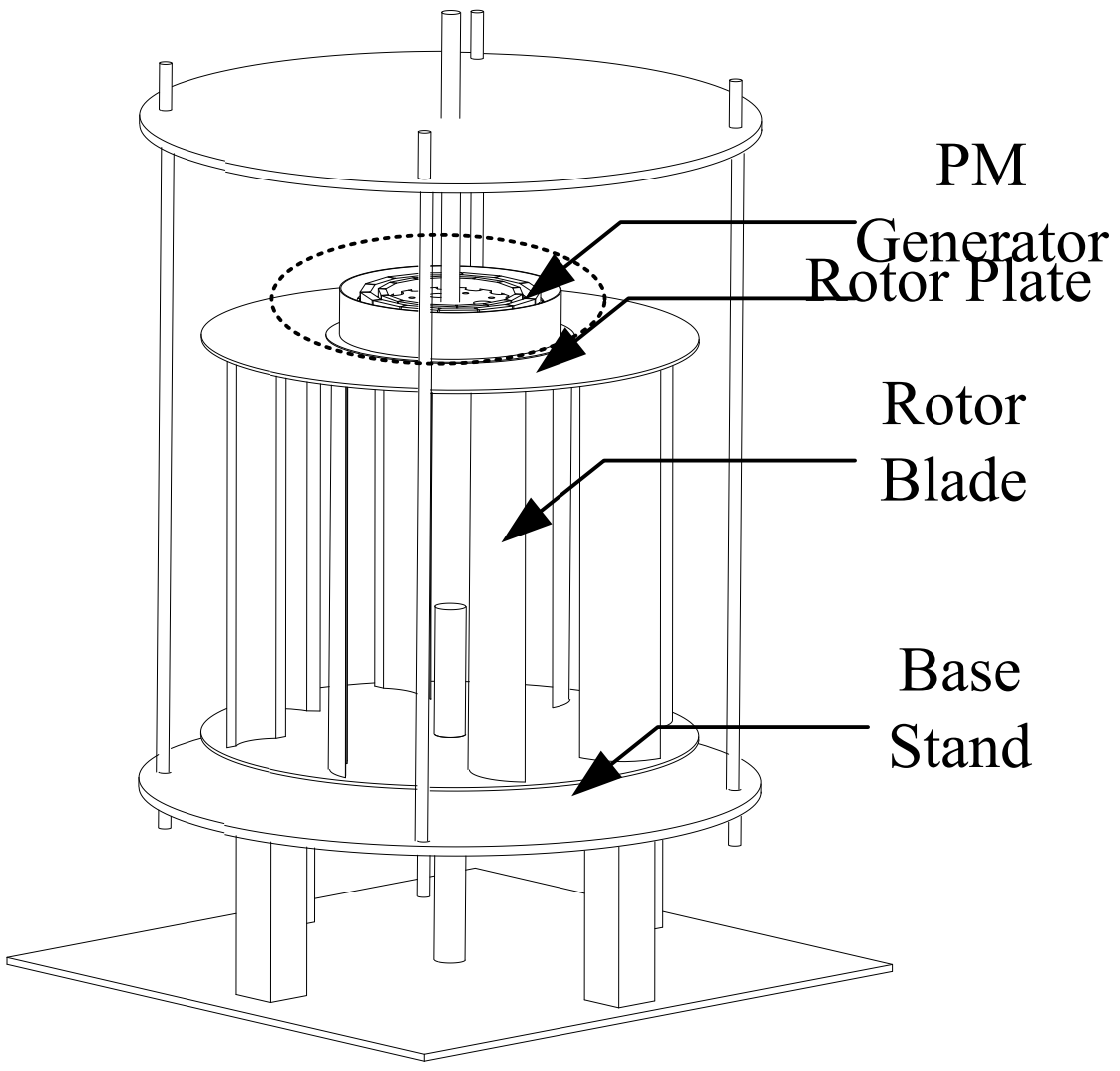

**Figure 3: Model design**

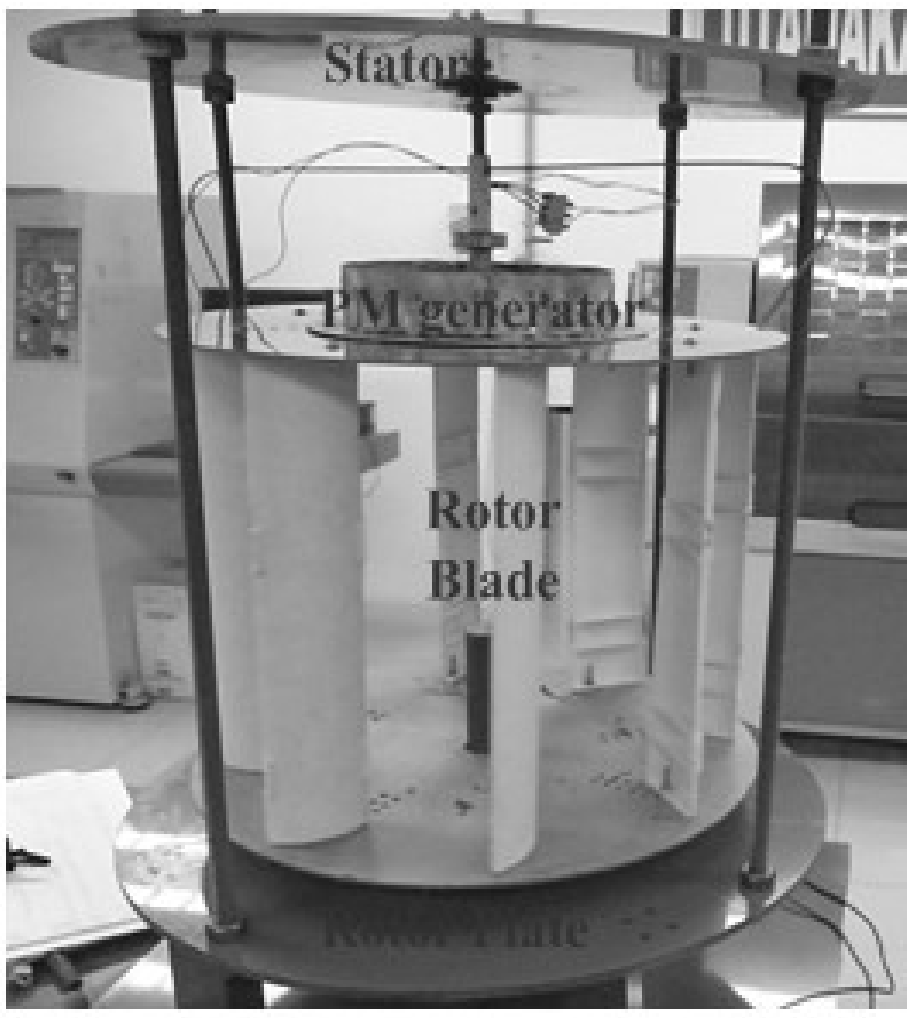

**Figure 4: Fabricated MAGLEV-VAWT**

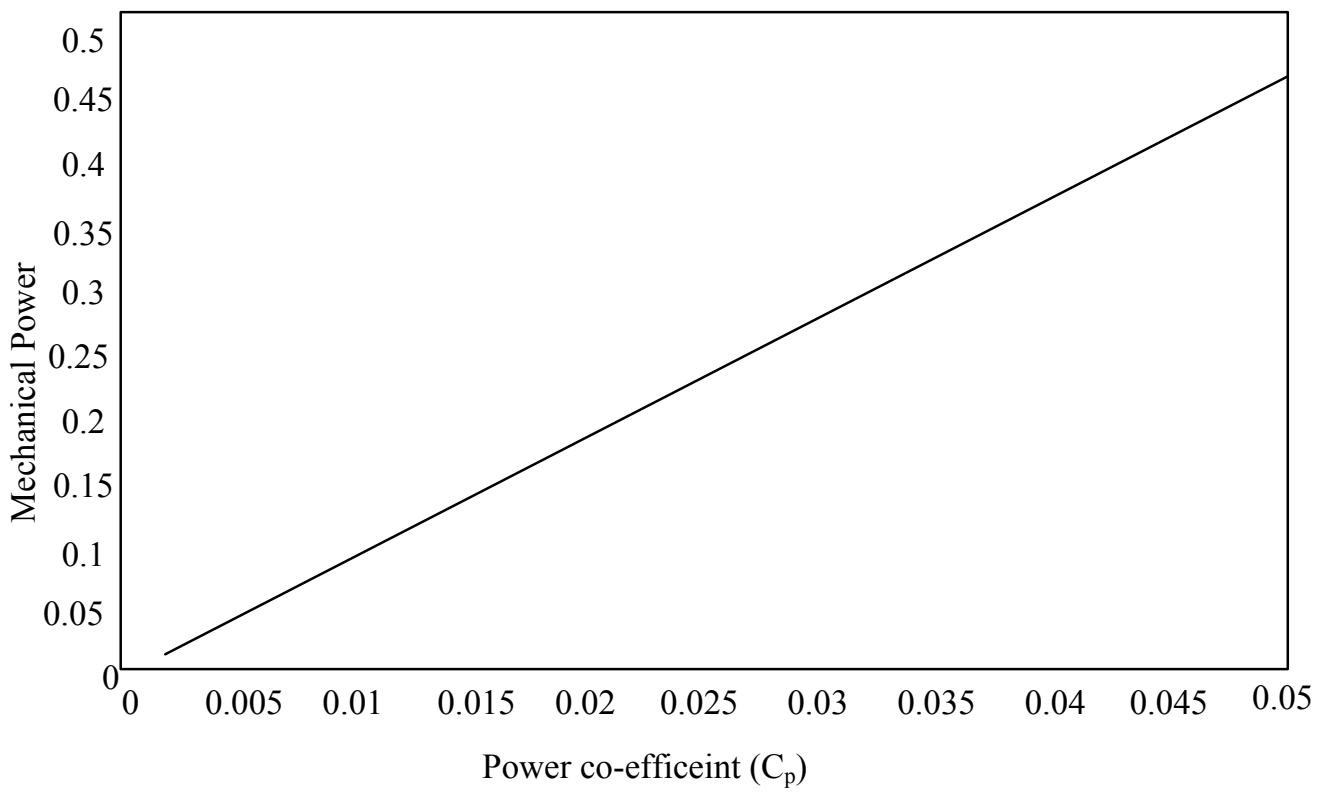

**Figure 5: Graph of $P_m$ Vs $C_p$ for current model**

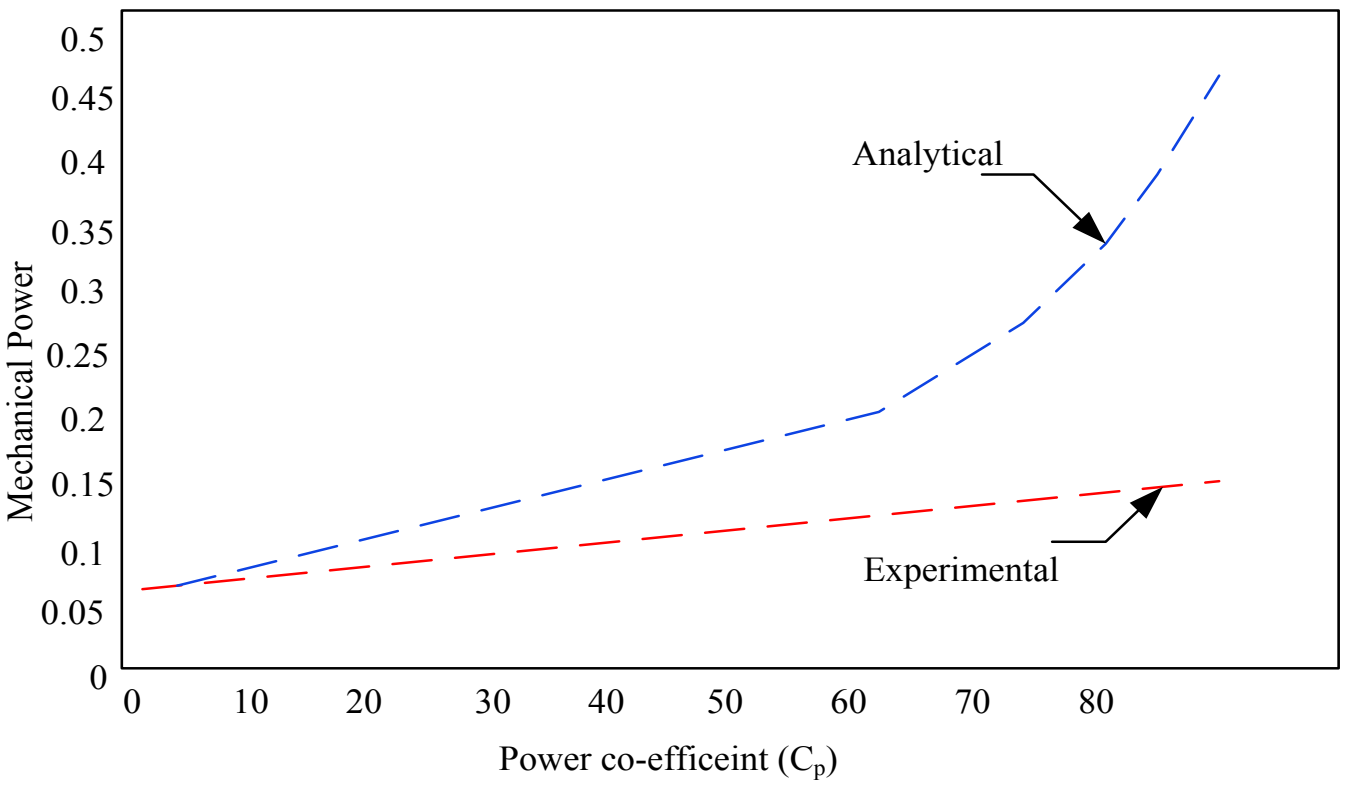

**Figure 6: Analytical Vs Experimental**

### *A. Area of the turbine*

From achieving the tip speed ratio, the radius of the turbine rotor is fixed. With the given value of the radius, and in achieving the power required, the height of the turbine can is fixed using Eq.(14)

$$h \uparrow = \frac{P_m}{R\downarrow \times V \times \rho \times C_p} \quad (14)$$

As seen for a given specific required mechanical power, an increase in the height of the turbine comes with a decrease in the radius of the turbine to achieve the same power and vice versa.

### *B. Number of blades*

In general, it is imperative that an increase in the number of blades should be better than fewer blades. An increase in the number of blades subsequently reduces the tip speed ratio of the turbine [3]. This therefore reduces the power coefficient leading the turbine to be less efficient than proposed. The increase in the number of blades subsequently increases the amount of drag faster than the amount of power generated. The number of blades required on the turbine for best efficiency is between 3 and 4 blades.

### *C. Turbine Speed*

The power generated from the turbine is further computed by using a variable speed directly driven generator like the PMSG. For the simulation therefore, the effect of varying the turbine speed is simulated and is seen to increase the power output of the turbine.

## IV. RESULTS AND DISCUSSION

The analysis is performed based on a required power output of 3500W at 250 RPM (26.18rad/s). The system behavior analysis is obtained by testing the design with different parameter characteristic. Parameters varied are the height, the radius, the turbine speed and the tip speed ratio. The designed toolbox help the designers to derive the closer dimensions for any rating of capacity but due considerations need to be given to the environmental and mechanical constraints. However, **Figure 7** shows the graphical output result from the designed toolbox and a comprehensive analysis is presented in **TABLE 1**. From the GUI result, to get the required speed, based on the calculations as in the previous section the radius of the turbine should be at about 0.5m while the minimum height is kept at about 5.8 m. With the constant value of the radius and rotor speed constant, the power of the wind turbine is seen to increase proportionally with respect to the increase in the height of the turbine rotor. Keeping the speed and the height constant, in order to achieve the rated power, the radius has to be around 0.62 m. Any further increase in the radius of the turbine leads to a subsequent decrease in the power of the turbine. The above be evidence for the relationship between the radius and height of the turbine. To achieve the required power, as the radius is increased, the height is seen to be decreased until the radius value is close to 0.7 m. At this point a subsequent increase in the radius leads to an increase in the height. This is because at 0.7m, the power coefficient is at its maximum and after this point; it begins to reduce again, leading to the need for more radius and height. **Figure 7** also shows the relationship between the power required and varying radius and height. From the figure graph, the best point for the required power is at the point when the height is about 1200m and the radius is about 0.6m.The power developed is increased as expected with the increase in the rotor speed and the relationship between the mechanical power and the tip speed ratio at a varying pitch angle. To achieve the required power, from the graph it shows that the tip speed ratio should be appeared less than 8.

The relationship between the torque and the turbine speed is also shown graphically with the radius and height of the rotor is kept at a constant value. This proves that the torque produced by the wind turbine increases with increase in mechanical speed until the maximum torque is achieved. A further increase in the speed then decreases the torque produced. This is because the mechanical speed affects the tip speed ratio and in turn affects the power coefficient. After the maximum speed is achieved, a subsequent increase in the speed increases the tip speed ratio to more than nine, thereby reducing the power coefficient.

## V. CONCLUSION

The analytical GUI system is designed and implemented using MATLAB software. However, It is found that system poses different behavior when the mechanical parameters are changed. In order to improve the turbine efficiency, the speed, the radius and the height of the turbine is to be adjusted accordingly. This modification can be explained as the following: Case 1: $R\uparrow W_m\downarrow H\downarrow$ Case 2: $R\downarrow W_m\uparrow H\uparrow$. This is to ensure that the tip speed ratio does not exceed or is not below the optimum tip speed ratio. With a variable speed generator the optimum way of designing the turbine rotor is to design the radius for the maximum speed of the generator and then design the height for the torque required.

**Figure 7.** Results from the GUI developed in MATLAB/SIMULINK

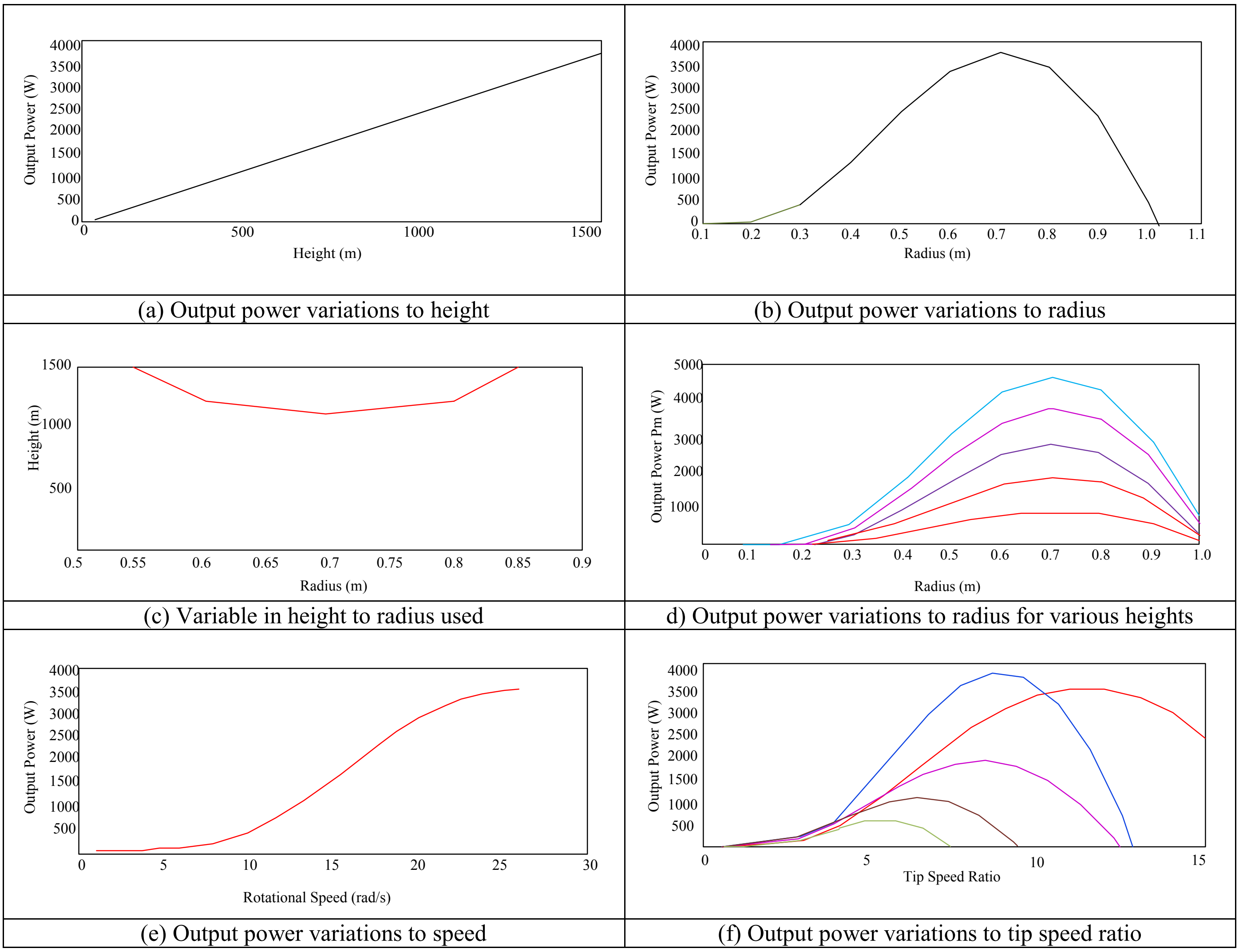

(a) Output power variations to height

(b) Output power variations to radius

(c) Variable in height to radius used

d) Output power variations to radius for various heights

(e) Output power variations to speed

(f) Output power variations to tip speed ratio

**TABLE 1. Parameters for various design values**

| Diameter (m) | Radius (m) | Speed (rpm) | Height (m) | Torque (N-m) |
|---|---|---|---|---|
| 1 | 0.5 | 286.4789 | 6.9107 | 12.2173 |
| 1.1 | 0.55 | 260.4354 | 6.2825 | 13.439 |
| 1.2 | 0.6 | 238.7324 | 5.7589 | 14.6608 |
| 1.3 | 0.65 | 220.3684 | 5.3159 | 15.8825 |
| 1.4 | 0.7 | 204.6278 | 4.9362 | 17.1042 |
| 1.5 | 0.75 | 190.9859 | 4.6071 | 18.326 |
| 1.6 | 0.8 | 179.0493 | 4.3192 | 19.5477 |
| 1.7 | 0.85 | 168.517 | 4.0651 | 20.7694 |